\documentclass{article}

\usepackage{arxiv}

\usepackage[utf8]{inputenc}
\usepackage[T1]{fontenc}
\usepackage{hyperref}
\usepackage{url}
\usepackage{booktabs}
\usepackage{amsfonts}
\usepackage{nicefrac}
\usepackage{microtype}
\usepackage{graphicx}
\usepackage{amsmath,amssymb}
\usepackage{bm}
\usepackage{amsthm}
\usepackage{algorithm}
\usepackage{algorithmicx}
\usepackage{algpseudocode}
\usepackage{multirow}
\usepackage{float}
\usepackage{subfig}
\usepackage[title]{appendix}

\title{Geometric Phase-Space Structure in Cosmological Solutions of Einstein's Field Equations}

\author{
  Hassan Ugail \\
  Centre for Visual Computing and Intelligent Systems \\
  University of Bradford \\
  United Kingdom \\
}

\begin{document}
\maketitle

\begin{abstract}
Einstein's field equations allow cosmological dynamics to depart from the Friedmann--Lema\^{i}tre--Robertson--Walker (FLRW) idealisation in several physically different ways. Matter may become spatially inhomogeneous, the local expansion scalar may vary across a hypersurface, the expansion may acquire anisotropic components through shear, and the free gravitational field may be encoded in nonzero Weyl curvature.  The key question is not only how far a model is from FLRW, but which geometric mechanism is responsible. A single departure-from-FLRW number cannot distinguish these mechanisms. This paper introduces a compact geometric diagnostic framework that keeps them separate while using standard quantities in general relativity. The framework is observer-explicit and domain-explicit, intended as a practical tool for comparing analytic and numerical solution families rather than as a new invariant classification of spacetime. Buchert's kinematical backreaction is retained as a derived explanatory quantity rather than a separate axis, since it is already fixed by the expansion-variance and shear contributions. A single curvature normalisation is used for all Weyl diagnostics. The method is tested on six benchmarks, namely FLRW, Bianchi-I, Kasner, Lema\^{i}tre--Tolman--Bondi dust, scalar-perturbed FLRW, and tensor-perturbed FLRW. These benchmarks occupy distinct regions of the diagnostic space, and the magnetic Weyl contribution appears only in the tensor case. The classification remains stable under changes of perturbation amplitude, spatial resolution, averaging domain, constraint reliability, and a leading-order observer tilt. The curvature expressions for the exact benchmarks are verified symbolically against metric-derived Weyl invariants, and the supporting computer code, numerical results, tables, and figures are publicly available.
\end{abstract}

\keywords{relativistic cosmology, inhomogeneity, Weyl curvature, cosmological backreaction, exact solutions, diagnostics}

\section{Introduction}\label{sec:intro}

The spatially homogeneous and isotropic Friedmann--Lema\^itre--Robertson--Walker (FLRW) models provide the standard reference class for relativistic cosmology. General solutions of Einstein's field equations, however, may depart from this reference in several physically distinct ways. Matter can become inhomogeneous, the local expansion rate can vary, the expansion can turn anisotropic, and the free gravitational field can develop electric or magnetic Weyl curvature. These mechanisms are not interchangeable. Two spacetimes may have the same value of a scalar departure measure while differing substantially in the geometric or dynamical origin of that departure. A diagnostic that compresses all deviations from FLRW behaviour into a single scalar can therefore obscure precisely the structure one wishes to identify.

A range of quantities is already used to characterise inhomogeneous relativistic cosmologies, including density contrast, shear, Weyl curvature scalars, Buchert's kinematical backreaction, and Hamiltonian or momentum constraint residuals \cite{buchert2000dust,buchert2001perfect,bck2011}. These diagnostics are individually informative, but they are not usually organised into a common representation that permits direct comparison between different exact, perturbative, and numerical solution families. This issue becomes particularly important in numerical-relativity cosmology, where simulations can generate large ensembles of inhomogeneous spacetimes whose qualitative behaviour must be summarised in a reproducible and interpretable way \cite{macpherson2017nr,aurrekoetxea2025review}.

The aim of this paper is not to introduce new curvature invariants. Rather, it is to organise standard kinematic and curvature quantities into a compact geometric diagnostic phase space for comparing cosmological dynamics. The contribution is methodological. First, we define dimensionless diagnostic axes that separately track matter inhomogeneity, expansion inhomogeneity, shear, and the electric and magnetic parts of the Weyl tensor, together with two reliability indicators associated with the Hamiltonian and momentum constraints. Second, we show that Buchert's kinematical backreaction is algebraically determined by the expansion-variance and shear contributions, and should therefore be treated as a derived explanatory quantity rather than as a separate diagnostic axis. Third, we validate the electric--magnetic Weyl split using a transverse-traceless gravitational-wave benchmark in which the magnetic Weyl contribution is genuinely nonzero. Fourth, we verify the analytic curvature expressions used for the Bianchi-I and Lema\^itre--Tolman--Bondi benchmarks by symbolic comparison with metric-derived Weyl invariants, and we demonstrate benchmark separation and robustness across exact and perturbative models. The broader diagnostic philosophy is to measure the preservation or loss of theoretically meaningful structure rather than to rely on a single generic score. Related ideas have been useful in other settings \cite{ugail2026qnn}, while the present construction is grounded entirely in relativistic geometry.

The resulting framework is observer-explicit, foliation-explicit, and domain-explicit. It is not proposed as a foliation-independent classification of spacetime. Instead, it is a transparent diagnostic defined relative to a specified observer field and averaging domain. The dependence on those choices is therefore not hidden, but is part of the physical specification of the comparison being made.

\section{Geometric diagnostic framework}\label{sec:framework}

\subsection{Observer kinematics and the electric and magnetic Weyl tensors}\label{sec:kin}

We work in the standard $1+3$ covariant description of relativistic cosmology \cite{ellisvanelst1999,emm2012,wald1984}. A timelike unit vector field $u^a$ with $u^a u_a=-1$ represents the family of fundamental observers, and $h_{ab}=g_{ab}+u_a u_b$ projects orthogonally to the flow. The covariant derivative of $u^a$ splits into the volume expansion $\theta=\nabla_a u^a$, the symmetric trace-free shear $\sigma_{ab}$, the vorticity, and the acceleration. The shear carries the departure from isotropic expansion, and its magnitude is measured through the scalar,
\begin{equation}
\sigma^2=\tfrac{1}{2}\,\sigma_{ab}\sigma^{ab}.
\end{equation}

The free gravitational field is encoded in the Weyl tensor $C_{abcd}$, which splits relative to $u^a$ into an electric part and a magnetic part,
\begin{equation}
E_{ab}=C_{acbd}\,u^c u^d,\qquad
B_{ab}=\tfrac{1}{2}\,\epsilon_{acde}\,C^{de}{}_{bf}\,u^c u^f,
\end{equation}
where $\epsilon_{acde}$ is the spacetime volume element. This split is the gravitational analogue of the decomposition of the electromagnetic field into electric and magnetic parts, and it carries a clear physical meaning. The electric part describes tidal forces, whereas the magnetic part is associated with frame dragging and with the transport of free gravitational energy by gravitational radiation \cite{costanatario2014,clifton2017}. We use the non-negative scalars $E_{ab}E^{ab}$ and $B_{ab}B^{ab}$ as the primary curvature quantities, and note for reference the relation,
\begin{equation}\label{eq:c2eb}
C_{abcd}C^{abcd}=8\,\big(E_{ab}E^{ab}-B_{ab}B^{ab}\big),
\end{equation}
which holds up to the sign convention of the curvature. Equation~\eqref{eq:c2eb} expresses the point that motivates the whole construction. A solution can carry a large magnetic Weyl field while $C_{abcd}C^{abcd}$ remains small or vanishes, so a diagnostic built on the scalar $C_{abcd}C^{abcd}$ alone is blind to purely radiative structure. Keeping $E_{ab}E^{ab}$ and $B_{ab}B^{ab}$ separate restores that information. Throughout, $B_{ab}$ denotes the magnetic part of the Weyl tensor and not a material magnetic field.

\subsection{Domain averages and the diagnostic axes}\label{sec:axes}

All diagnostics are defined as domain averages on the chosen spatial hypersurface. For a scalar field $X$ and a domain $\mathcal D$ we write,
\begin{equation}
\langle X\rangle_{\mathcal D}=\frac{\int_{\mathcal D} X\,\mathrm{d}V}{\int_{\mathcal D}\mathrm{d}V},
\end{equation}
with $\mathrm{d}V$ the proper volume element induced on the hypersurface. The diagnostic axes are then dimensionless ratios that compare a measure of structure with a suitable mean or curvature scale. We define,
\begin{equation}
I_\rho=\frac{\mathrm{Var}_{\mathcal D}(\rho)}{\langle\rho\rangle_{\mathcal D}^2+\epsilon},\qquad
I_\theta=\frac{\mathrm{Var}_{\mathcal D}(\theta)}{\langle\theta\rangle_{\mathcal D}^2+\epsilon},\qquad
I_\sigma=\frac{\langle\sigma^2\rangle_{\mathcal D}}{\langle\theta\rangle_{\mathcal D}^2+\epsilon},
\end{equation}
for the matter inhomogeneity, the expansion inhomogeneity, and the shear, and,
\begin{equation}
I_E=\frac{\langle E_{ab}E^{ab}\rangle_{\mathcal D}}{K_{\mathcal D}+\epsilon},\qquad
I_B=\frac{\langle B_{ab}B^{ab}\rangle_{\mathcal D}}{K_{\mathcal D}+\epsilon},
\end{equation}
for the electric and magnetic Weyl curvature. Here $\epsilon$ is a small regulator that keeps the ratios finite in the homogeneous limit, $\mathrm{Var}_{\mathcal D}(X)=\langle X^2\rangle_{\mathcal D}-\langle X\rangle_{\mathcal D}^2$, and $K_{\mathcal D}$ is a curvature normalisation discussed below. Two further indicators $I_{\mathcal H}$ and $I_{\mathcal M}$ measure the violation of the Hamiltonian and momentum constraints and serve as reliability flags rather than as physical structure. The full diagnostic vector is,
\begin{equation}
\mathbf{X}_{\mathrm{cosmo}}(\mathcal D,t)=\big(I_\rho,\,I_\theta,\,I_\sigma,\,I_E,\,I_B,\,I_{\mathcal H},\,I_{\mathcal M}\big).
\end{equation}
The term non-redundant is used here in an algebraic sense. No Buchert-type derived coordinate is carried as a separate axis, as is shown for $Q_{\mathcal D}$ below. We do not claim that the axes are dynamically independent. The Einstein equations couple them through the constraint and propagation equations, and the principal component analysis of Section~\ref{sec:results} quantifies the resulting correlations across the benchmark suite.

The curvature scale $K_{\mathcal D}$ must be specified explicitly, since comparing $I_E$ and $I_B$ across solution families is only meaningful when the same scale is used throughout. We adopt a single normalisation for every benchmark, with the domain-averaged fourth power of the expansion as,
\begin{equation}\label{eq:Kbg}
K_{\mathcal D}=\langle\theta^4\rangle_{\mathcal D}+\epsilon.
\end{equation}
This choice is positive for matter, vacuum, and radiative models alike, and it introduces no discontinuity for a model that evolves from one regime to another. An alternative is the reference FLRW Kretschmann scale $12[(\dot H+H^2)^2+H^4]$ built from the mean expansion $H=\langle\theta\rangle_{\mathcal D}/3$, which is retained only for the sensitivity check as discussed in Section~\ref{sec:robust}. The qualitative separation of the benchmarks, and in particular the dominant axis of each, is unchanged between the two choices.

\subsection{Backreaction as a derived quantity and diagnostic labels}\label{sec:derived}

The Buchert kinematical backreaction for an irrotational dust domain is,
\begin{equation}\label{eq:Q}
Q_{\mathcal D}=\tfrac{2}{3}\big(\langle\theta^2\rangle_{\mathcal D}-\langle\theta\rangle_{\mathcal D}^2\big)-2\langle\sigma^2\rangle_{\mathcal D},
\end{equation}
which combines the variance of the expansion with the mean shear \cite{buchert2000dust,buchert2001perfect}. $Q_{\mathcal D}$ is often treated as a primary measure of inhomogeneity, yet its normalised magnitude is not independent of the axes already introduced. Dividing Equation~\eqref{eq:Q} by $\langle\theta\rangle_{\mathcal D}^2$ gives,
\begin{equation}\label{eq:Qid}
\frac{|Q_{\mathcal D}|}{\langle\theta\rangle_{\mathcal D}^2}=\Big|\tfrac{2}{3}\,I_\theta-2\,I_\sigma\Big|,
\end{equation}
so the backreaction amplitude is fixed algebraically once $I_\theta$ and $I_\sigma$ are known. We therefore retain $Q_{\mathcal D}$ as a derived explanatory quantity, useful for distinguishing variance-driven from shear-driven backreaction, but it does not warrant treatment as a separate axis. The same reasoning shows that a naive vector including a separate backreaction axis would in fact be of lower dimension than its component count suggests.

For convenience, each diagnostic vector is also assigned a descriptive label according to its dominant axis, subject to a small floor below which a state is called FLRW-like and a reliability tolerance above which it is flagged as numerically unreliable. These labels summarise the component vector and nothing more. The primary output is the component vector itself, together with its trajectory through the diagnostic phase space, and the classifier-free separability of the benchmark families is examined directly (as discussed in Section~\ref{sec:results}).

\section{Benchmark cosmologies}\label{sec:benchmarks}

The framework is exercised on six benchmarks, each chosen to isolate a different mechanism of departure from FLRW behaviour. Table~\ref{tab:models} summarises the models and the signatures they are expected to produce.

\subsection{Exact homogeneous and vacuum models}\label{sec:homog}

The flat FLRW model serves as the reference case, for which every structure axis vanishes identically. The Bianchi-I model describes homogeneous but anisotropic expansion through the metric,
\begin{equation}
\mathrm{d}s^2=-\mathrm{d}t^2+a_x^2(t)\,\mathrm{d}x^2+a_y^2(t)\,\mathrm{d}y^2+a_z^2(t)\,\mathrm{d}z^2,
\end{equation}
with power-law scale factors $a_i(t)=t^{p_i}$. Because the model is spatially homogeneous, the matter and expansion variances vanish, while the shear and electric Weyl curvature remain nonzero. The magnetic Weyl part vanishes identically for the comoving congruence, so Bianchi-I is purely electric. The Kasner solution is the vacuum member of this family, defined by the conditions,
\begin{equation}
p_1+p_2+p_3=1,\qquad p_1^2+p_2^2+p_3^2=1.
\end{equation}
This provides a strongly anisotropic case with no matter content \cite{kasner1921,wainwrightkrasinski2008,misner1969}. Because Kasner carries no density, it is normalised by the vacuum expansion scale rather than a dust scale, which confirms that the framework does not require the presence of matter.

\subsection{Exact inhomogeneous dust}\label{sec:ltb}

The Lema\^itre--Tolman--Bondi class describes spherically symmetric inhomogeneous dust and supplies an exact inhomogeneous benchmark with nonzero electric Weyl curvature \cite{lemaitre1933,tolman1934,bondi1947,krasinski1997,bck2011}. In the parabolic case, the areal radius $R(t,r)$ obeys,
\begin{equation}
\dot R^2=\frac{2M(r)}{R},
\end{equation}
with $M(r)$ the mass function, the rest-mass density is,
\begin{equation}
\rho=\frac{M'}{4\pi R^2 R'},
\end{equation}
and the Weyl invariant takes the closed form,
\begin{equation}\label{eq:ltbweyl}
C_{abcd}C^{abcd}=48\left(\frac{M}{R^3}-\frac{M'}{3R^2 R'}\right)^2,
\end{equation}
where a prime denotes a radial derivative. The model produces matter inhomogeneity, expansion inhomogeneity, shear, and electric Weyl curvature together. The magnetic Weyl part again vanishes for the comoving congruence.

\subsection{Constraint-consistent perturbations}\label{sec:perturb}

Two perturbative benchmarks complete the set. The scalar case uses the longitudinal gauge,
\begin{equation}
\mathrm{d}s^2=a^2(\eta)\big[-(1+2\Phi)\,\mathrm{d}\eta^2+(1-2\Phi)\,\delta_{ij}\,\mathrm{d}x^i\mathrm{d}x^j\big],
\end{equation}
with $\Phi$ the metric potential and $\eta$ conformal time. The density and velocity perturbations are constructed from the linearised Hamiltonian and momentum constraints for a matter-dominated growing mode, so that the benchmark is a genuine solution of the linearised field equations rather than a prescribed test field \cite{bardeen1980,kodamasasaki1984,mfb1992}. The tensor case uses a transverse-traceless perturbation,
\begin{equation}
\mathrm{d}s^2=a^2(\eta)\big[-\mathrm{d}\eta^2+(\delta_{ij}+h^{\mathrm{TT}}_{ij})\,\mathrm{d}x^i\mathrm{d}x^j\big],
\end{equation}
which describes a propagating gravitational wave on a radiation-dominated FLRW background. The two perturbative benchmarks therefore use different backgrounds, matter-dominated for the scalar mode and radiation-dominated for the tensor mode, which is the natural setting for each. A tensor mode carries both electric and magnetic Weyl components, so this benchmark activates the magnetic-Weyl axis.

Each diagnostic is evaluated in the order of the linearised metric and constraints used to construct the benchmark. For the scalar case, the shear axis $I_\sigma$ is computed from the linearised metric and is identically zero by construction, since scalar-induced shear first appears at second order in the perturbation amplitude and is not included here. The scalar signature therefore reflects this linear truncation, with matter inhomogeneity as the leading axis. For the tensor case, the electric and magnetic invariants are equal only for a plane wave in flat space. On the expanding background, the mode carries a finite ratio of wavelength to Hubble scale, so $E_{ab}E^{ab}$ and $B_{ab}B^{ab}$ are comparable rather than exactly equal, and the Weyl scalar $C_{abcd}C^{abcd}=8(E_{ab}E^{ab}-B_{ab}B^{ab})$ is suppressed relative to either part while $B_{ab}B^{ab}$ remains nonzero. In fully nonlinear general relativity $C_{abcd}C^{abcd}$ is not generally small, so this suppression is a property of the linear radiative mode rather than a general statement.

\begin{table}[t]
\caption{Benchmark cosmologies and the diagnostic signatures they are expected to produce. The entry $I_B=0$ for the first five models refers to the comoving congruence, for which those models are purely electric.}\label{tab:models}
\begin{tabular}{@{}lll@{}}
\toprule
Model & Character & Expected signature\\
\midrule
FLRW & isotropic baseline & all axes vanish\\
Bianchi-I & homogeneous anisotropic & $I_\sigma>0,\ I_E>0,\ I_B=0$\\
Kasner & vacuum anisotropic & $I_\rho=0,\ I_\sigma>0,\ I_E>0,\ I_B=0$\\
LTB dust & inhomogeneous dust & $I_\rho>0,\ I_E>0,\ I_B=0$\\
Scalar-perturbed FLRW & matter inhomogeneity & $I_\rho>0,\ I_B=0,\ I_{\mathcal H},I_{\mathcal M}\!\approx\!0$\\
Tensor-perturbed FLRW & gravitational wave & $I_E>0,\ I_B>0$\\
\bottomrule
\end{tabular}
\end{table}

\section{Verification and consistency}\label{sec:verification}

Before presenting the results, we verify that the analytic ingredients are correct. A diagnostic based on an incorrect curvature expression would be unreliable, irrespective of the apparent clarity of its numerical output. The curvature expressions for the Bianchi-I and LTB benchmarks were derived independently from the metric by symbolic computation, by forming the Christoffel symbols, the Riemann and Ricci tensors, and the Weyl tensor, and contracting to obtain $C_{abcd}C^{abcd}$. For Bianchi-I, the symbolic invariant and the closed form agree as an exact identity, the invariant vanishes in the isotropic limit, and the magnetic Weyl tensor is identically zero. For LTB, the closed form in Equation~\eqref{eq:ltbweyl} agrees with the metric-derived invariant to machine precision across a grid of points, the invariant vanishes in the homogeneous limit, and the parabolic solution satisfies its field equation exactly. The same machinery reproduces the known vacuum Kasner value of the Weyl invariant, which provides a further check.

Two further checks concern the internal consistency of the framework. The backreaction identity of Equation~\eqref{eq:Qid} was confirmed numerically, with the difference between the directly computed $|Q_{\mathcal D}|/\langle\theta\rangle_{\mathcal D}^2$ and the reconstruction from $I_\theta$ and $I_\sigma$ lying at the level of floating-point round-off. The linearised Hamiltonian and momentum residuals of the scalar benchmark likewise lie at round-off, which certifies that the benchmark is a genuine linear solution. The magnetic Weyl axis was checked directly. It vanishes to round-off for the five non-radiative benchmarks and remains nonzero for the tensor benchmark, where the electric and magnetic parts are comparable, with a residual $|I_E-I_B|/\max(I_E,I_B)\approx0.13$ that decreases as the mode becomes more sub-horizon. Table~\ref{tab:verify} shows these results.

\begin{table}[t]
\caption{Verification and consistency checks. Residuals are reported as maximum values over the relevant grid or time range.}\label{tab:verify}
\begin{tabular}{@{}ll@{}}
\toprule
Check & Result\\
\midrule
Bianchi-I Weyl, symbolic identity & $C^2_{\mathrm{metric}}-C^2_{\mathrm{closed}}=0$ (exact)\\
Bianchi-I isotropic limit & $C_{abcd}C^{abcd}=0$\\
Bianchi-I magnetic Weyl & $B_{ab}=0$\\
Bianchi-I anisotropy sweep & metric vs closed form $\sim1.2\times10^{-12}$\\
LTB Weyl, metric vs closed form & $0$ (machine precision)\\
LTB homogeneous limit & $C_{abcd}C^{abcd}=0$\\
LTB field equation & $\dot R^2-2M/R=0$ (exact)\\
Backreaction identity & $\big||Q_{\mathcal D}|/\langle\theta\rangle^2-|\tfrac{2}{3}I_\theta-2I_\sigma|\big|\sim1.1\times10^{-15}$\\
Scalar perturbation, Hamiltonian & $I_{\mathcal H}\sim1.1\times10^{-16}$\\
Scalar perturbation, momentum & $I_{\mathcal M}\sim6.9\times10^{-18}$\\
Magnetic axis, non-radiative benchmarks & $I_B=0$ (round-off)\\
Magnetic axis, tensor benchmark & $I_B>0$, $|I_E-I_B|/\max\approx0.13$\\
\bottomrule
\end{tabular}
\end{table}

\section{Results}\label{sec:results}

\begin{figure}[t]
\centering
\includegraphics[width=0.92\textwidth]{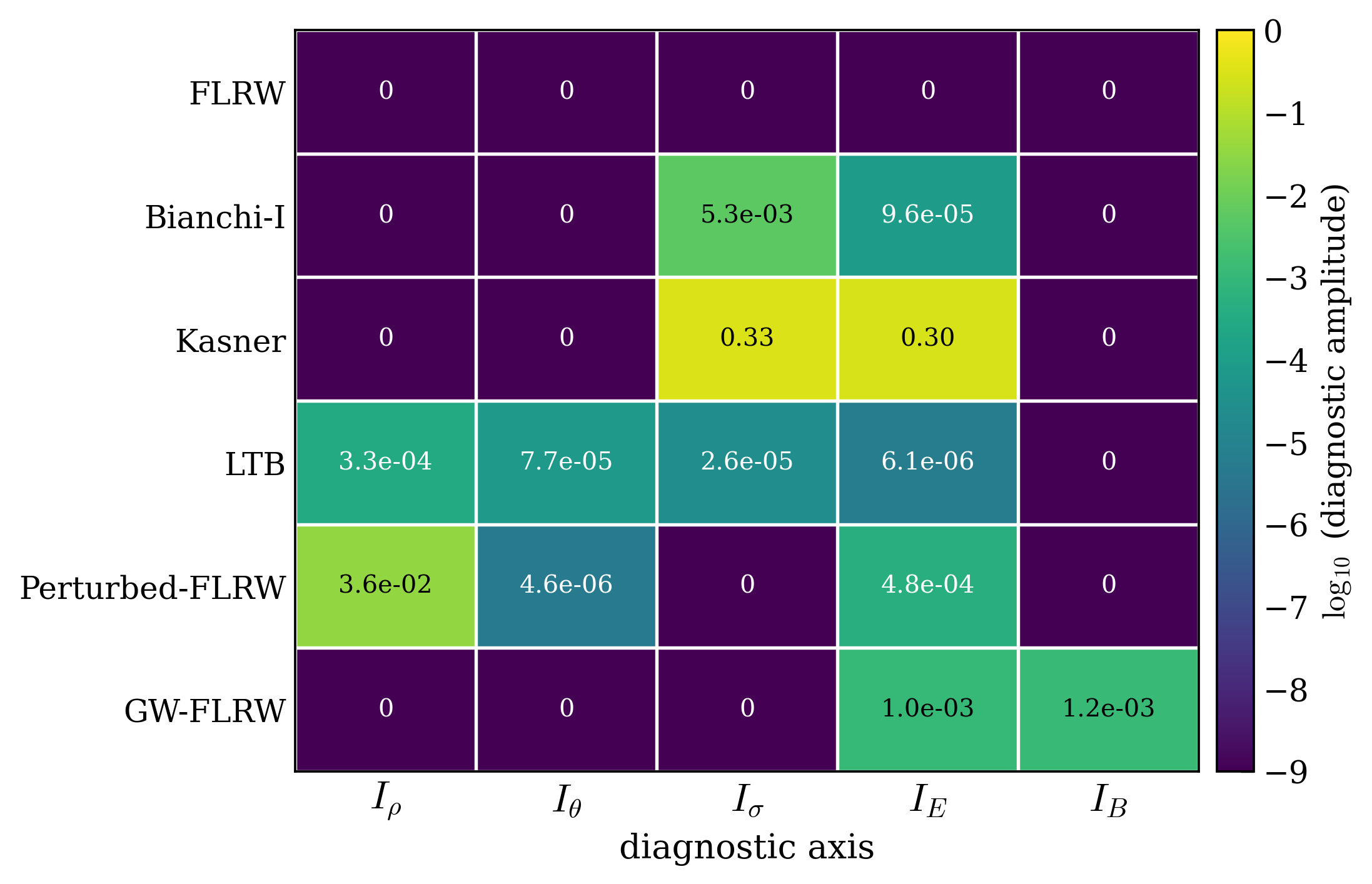}
\caption{Final-time diagnostic profile for the six benchmarks. Cells show the dimensionless axis values on a logarithmic colour scale, with entries below the round-off floor shown as zero. Each model occupies a distinct signature pattern, and the magnetic Weyl axis $I_B$ is populated only by the gravitational-wave benchmark.}\label{fig:heatmap}
\end{figure}

The diagnostic vector was computed along the dynamical history of each benchmark. Figure~\ref{fig:heatmap} shows the final-time profile across all six models on a logarithmic colour scale, and Table~\ref{tab:final} reports the same values together with the dominant axis. The structure of these results constitutes the central finding of the paper. Each benchmark family occupies a distinct region of the diagnostic space, and the region matches the physical mechanism that defines the model. The FLRW model lies at the origin, with every axis at round-off. Bianchi-I and Kasner are dominated by shear and electric Weyl curvature, with Kasner an order of magnitude stronger because it is a strongly anisotropic vacuum solution rather than a mild homogeneous anisotropy. The LTB model distributes a moderate signal across matter inhomogeneity, expansion inhomogeneity, shear, and electric Weyl curvature, as expected for an exact inhomogeneous dust spacetime. The scalar-perturbed model is dominated by matter inhomogeneity. The gravitational-wave benchmark is the only model with a nonzero magnetic Weyl axis.

\begin{table}[t]
\caption{Final-time diagnostic values and dominant axis. Values below $10^{-12}$ are at the round-off floor and are written as zero.}\label{tab:final}
\begin{tabular}{@{}lcccccl@{}}
\toprule
Model & $I_\rho$ & $I_\theta$ & $I_\sigma$ & $I_E$ & $I_B$ & Dominant\\
\midrule
FLRW & $0$ & $0$ & $0$ & $0$ & $0$ & none\\
Bianchi-I & $0$ & $0$ & $5.3\times10^{-3}$ & $9.6\times10^{-5}$ & $0$ & $I_\sigma$\\
Kasner & $0$ & $0$ & $0.333$ & $0.296$ & $0$ & $I_\sigma$\\
LTB & $3.3\times10^{-4}$ & $7.7\times10^{-5}$ & $2.6\times10^{-5}$ & $6.1\times10^{-6}$ & $0$ & $I_\rho$\\
Perturbed-FLRW & $3.6\times10^{-2}$ & $4.6\times10^{-6}$ & $0$ & $4.8\times10^{-4}$ & $0$ & $I_\rho$\\
GW-FLRW & $0$ & $0$ & $0$ & $1.0\times10^{-3}$ & $1.2\times10^{-3}$ & $I_B$\\
\bottomrule
\end{tabular}
\end{table}

\begin{figure}[t]
\centering
\includegraphics[width=\textwidth]{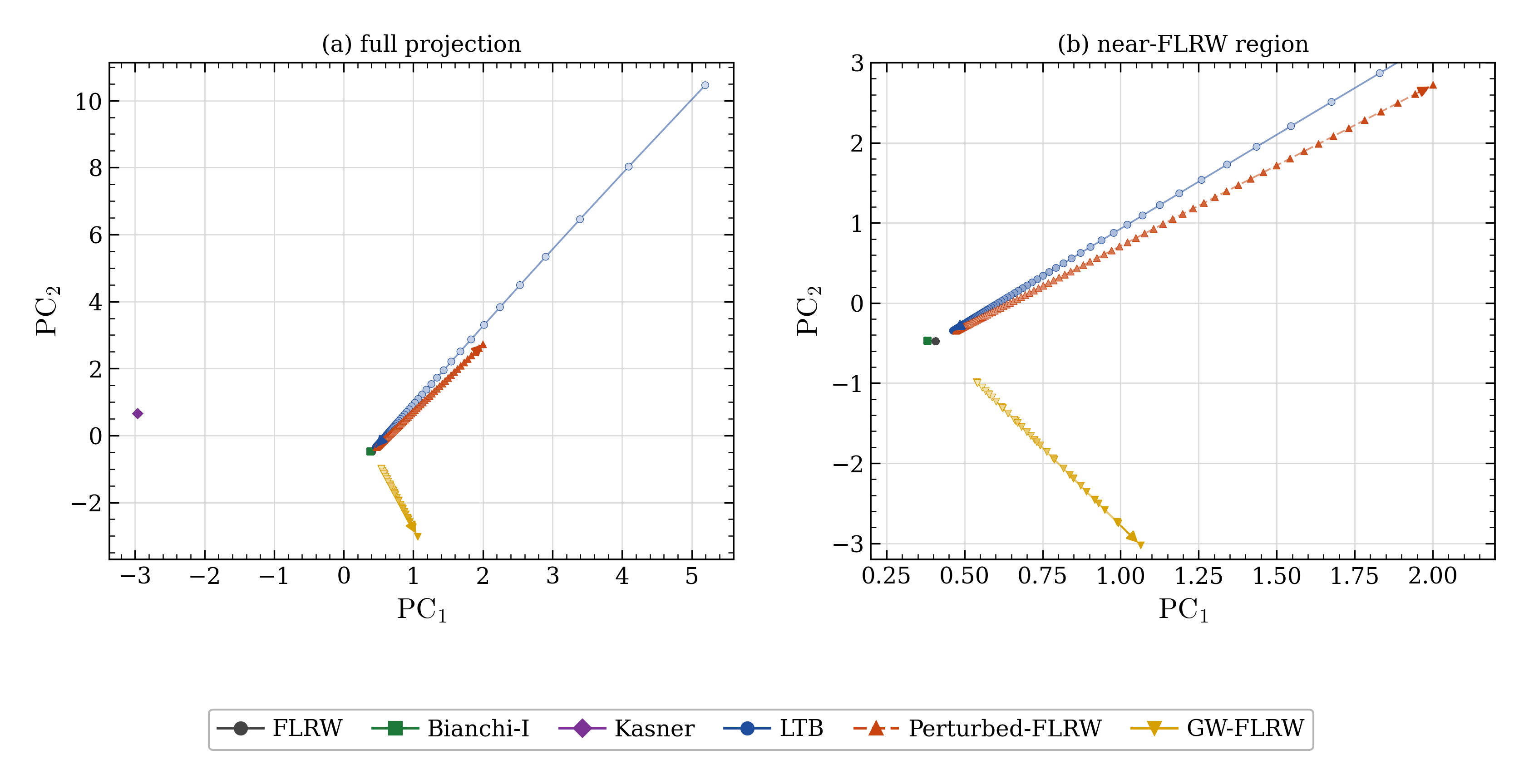}
\caption{Principal component projection of the diagnostic vectors across the benchmark histories. Panel (a) shows the full projection, including the extended LTB branch and the isolated Kasner point. Panel (b) zooms into the near-FLRW region, where initially neighbouring trajectories separate into distinct branches. The upper part of the LTB branch is intentionally outside the zoomed range. The projection is used as a visual separability check rather than as an unsupervised classifier.}\label{fig:pca}
\end{figure}

The separation of the families can be examined without reference to the descriptive labels. A principal component projection of the diagnostic vectors, shown in Figure~\ref{fig:pca}, places the families on visibly distinct branches, with the first two components accounting for $71.3\%$ of the variance. The separation is quantified in a label-free way by measuring how well the diagnostic vector alone identifies the physical regime. A silhouette score computed against the known regimes rises from about $0.47$ over the full history to about $0.78$ in the developed late-time regime, and a leave-one-out nearest-centroid assignment, in which each diagnostic vector is assigned to the regime whose mean it lies closest to while being held out of that mean, recovers the correct regime with an accuracy of about $0.80$ over the full history and about $0.96$ at late times. The full-history figures are lower for a physical reason rather than a technical one, since at early times every trajectory lies close to the FLRW point and the families have not yet separated. We therefore present the projection as a visual separability check and report the developed-regime separability as the quantitative statement, rather than relying on an unsupervised clustering label. Those two components account for most of the variance, which itself is informative, since it reflects the dynamical coupling of the axes through the Einstein equations rather than any redundancy in their definitions, the one genuine redundancy having already been removed by treating $Q_{\mathcal D}$ as derived.

\begin{figure}[t]
\centering
\includegraphics[width=0.82\textwidth]{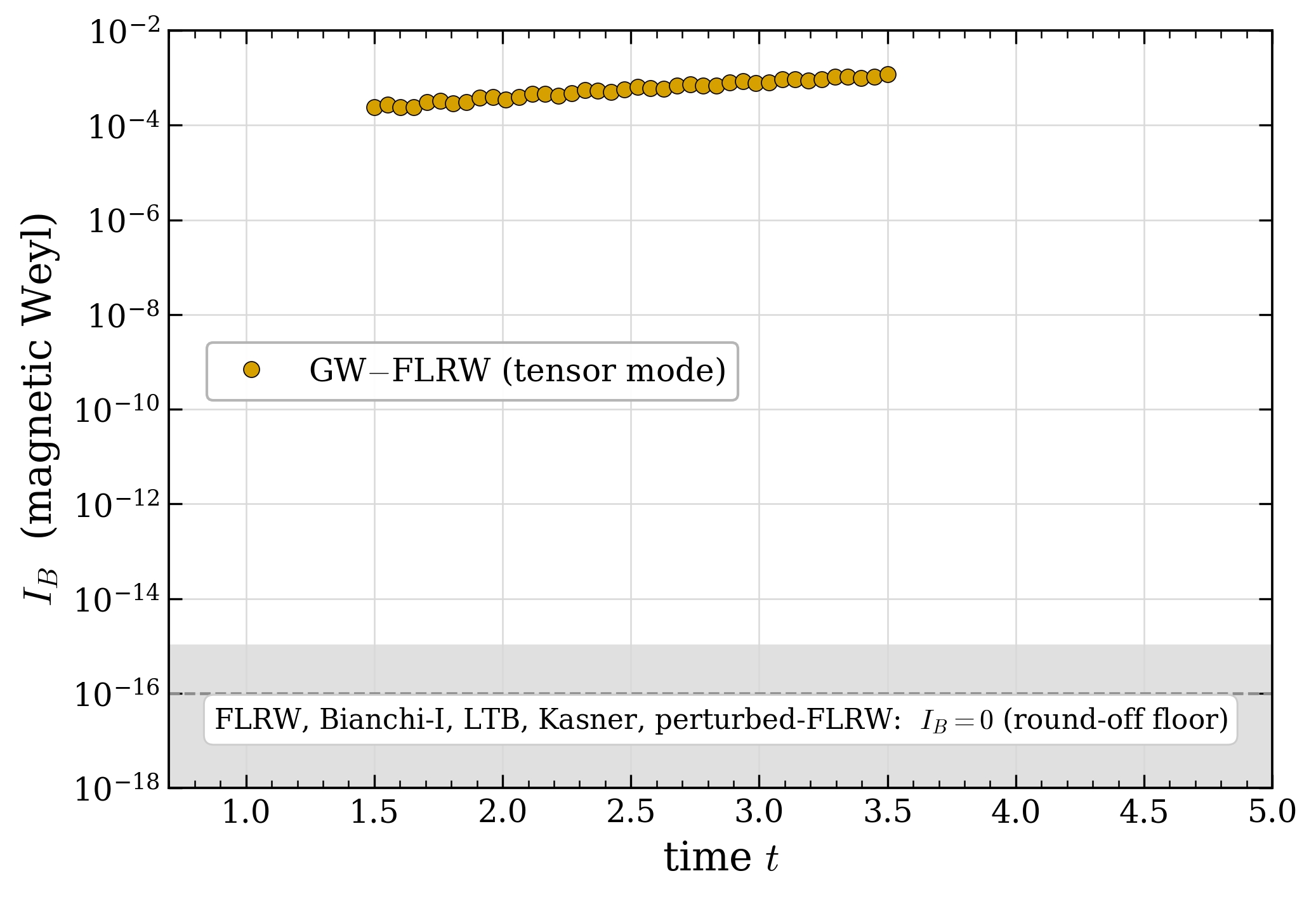}
\caption{Magnetic Weyl axis $I_B$ as a function of time. The gravitational-wave benchmark carries a clearly nonzero magnetic Weyl amplitude, while FLRW, Bianchi-I, LTB, Kasner, and the scalar-perturbed model remain at the round-off floor, consistent with their purely electric character.}\label{fig:magnetic}
\end{figure}

The electric and magnetic Weyl split is the most distinctive feature of the framework, and the gravitational-wave benchmark demonstrates its value directly. Figure~\ref{fig:magnetic} shows the magnetic axis as a function of time for all six models. The five non-radiative benchmarks return $I_B$ at the round-off floor, consistent with their purely electric character, whereas the tensor benchmark carries a magnetic Weyl amplitude many orders of magnitude larger. For this sub-horizon mode, the electric and magnetic amplitudes are comparable, so the Weyl scalar $C_{abcd}C^{abcd}=8(E_{ab}E^{ab}-B_{ab}B^{ab})$ is small and even changes sign. This is precisely the situation in which a scalar Weyl diagnostic would report almost nothing while the magnetic axis reports a clear signal. In cosmological terms, this reflects the fact that many standard solutions force the magnetic Weyl part to vanish, so that its appearance is a marker of genuinely radiative or frame-dragging structure \cite{clifton2017,costanatario2014}.

The derived backreaction completes the picture. Decomposing $Q_{\mathcal D}$ into its variance and shear contributions shows that the backreaction of the homogeneous anisotropic and vacuum benchmarks is shear-driven, whereas in the inhomogeneous dust case the expansion-variance contribution becomes relevant. This interpretation follows directly from the values of $I_\theta$ and $I_\sigma$ through Equation~\eqref{eq:Qid}, which is precisely why the backreaction does not require a coordinate of its own.

\section{Robustness and sensitivity}\label{sec:robust}

\begin{figure}[t]
\centering
\subfloat[Bianchi-I anisotropy sweep]{\includegraphics[width=0.33\textwidth]{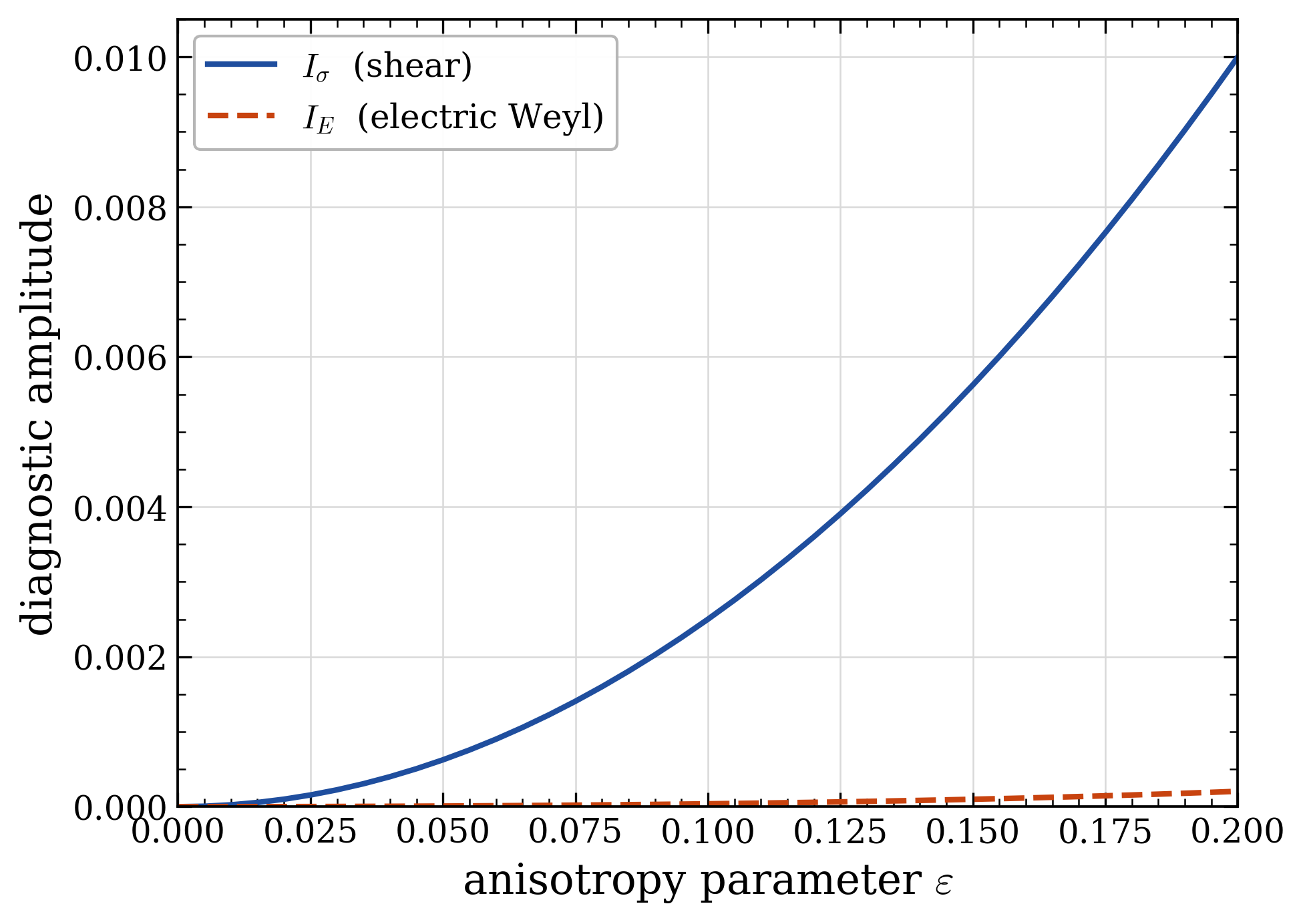}}\hfill
\subfloat[LTB amplitude sweep]{\includegraphics[width=0.33\textwidth]{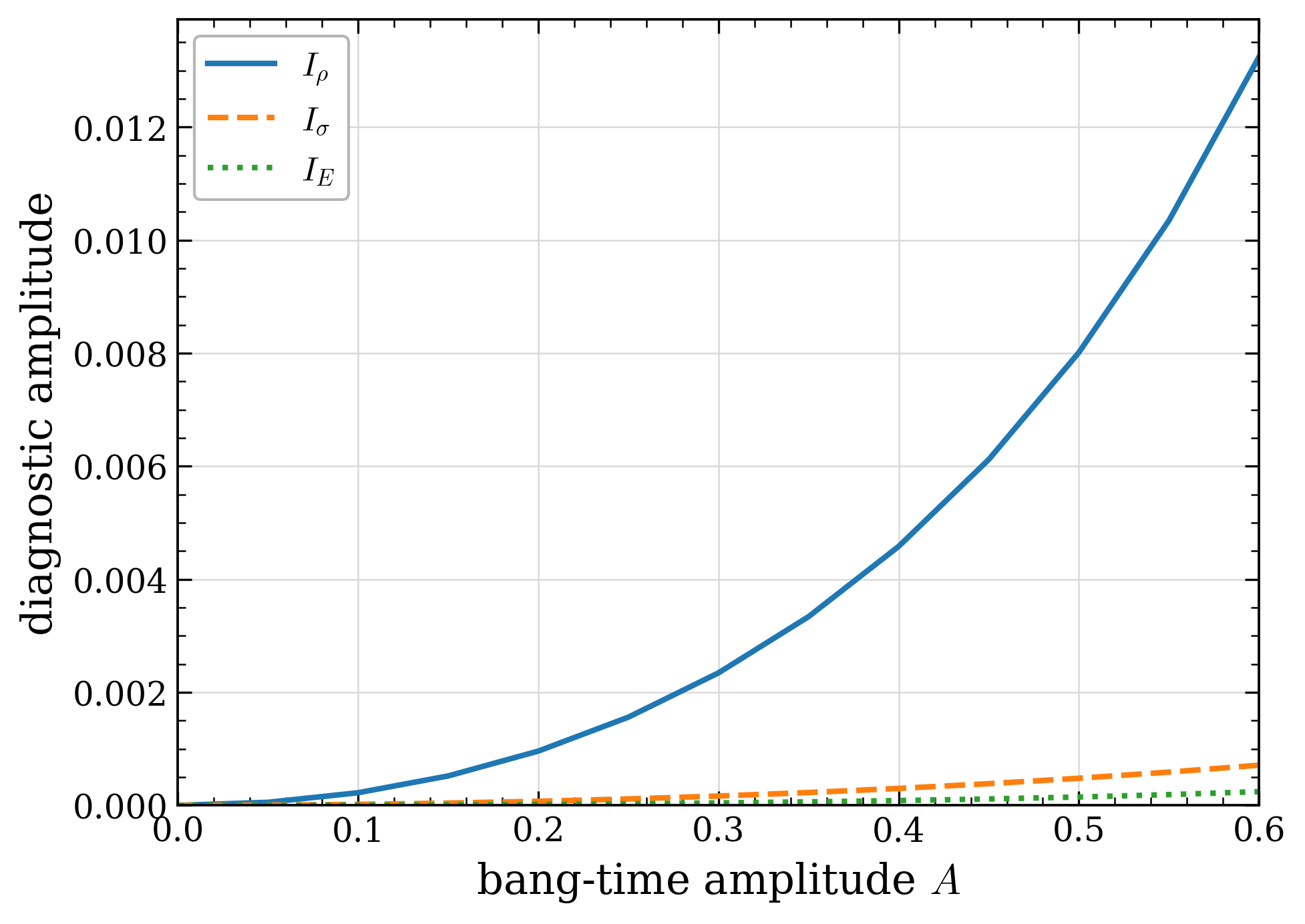}}\hfill
\subfloat[LTB radial resolution]{\includegraphics[width=0.33\textwidth]{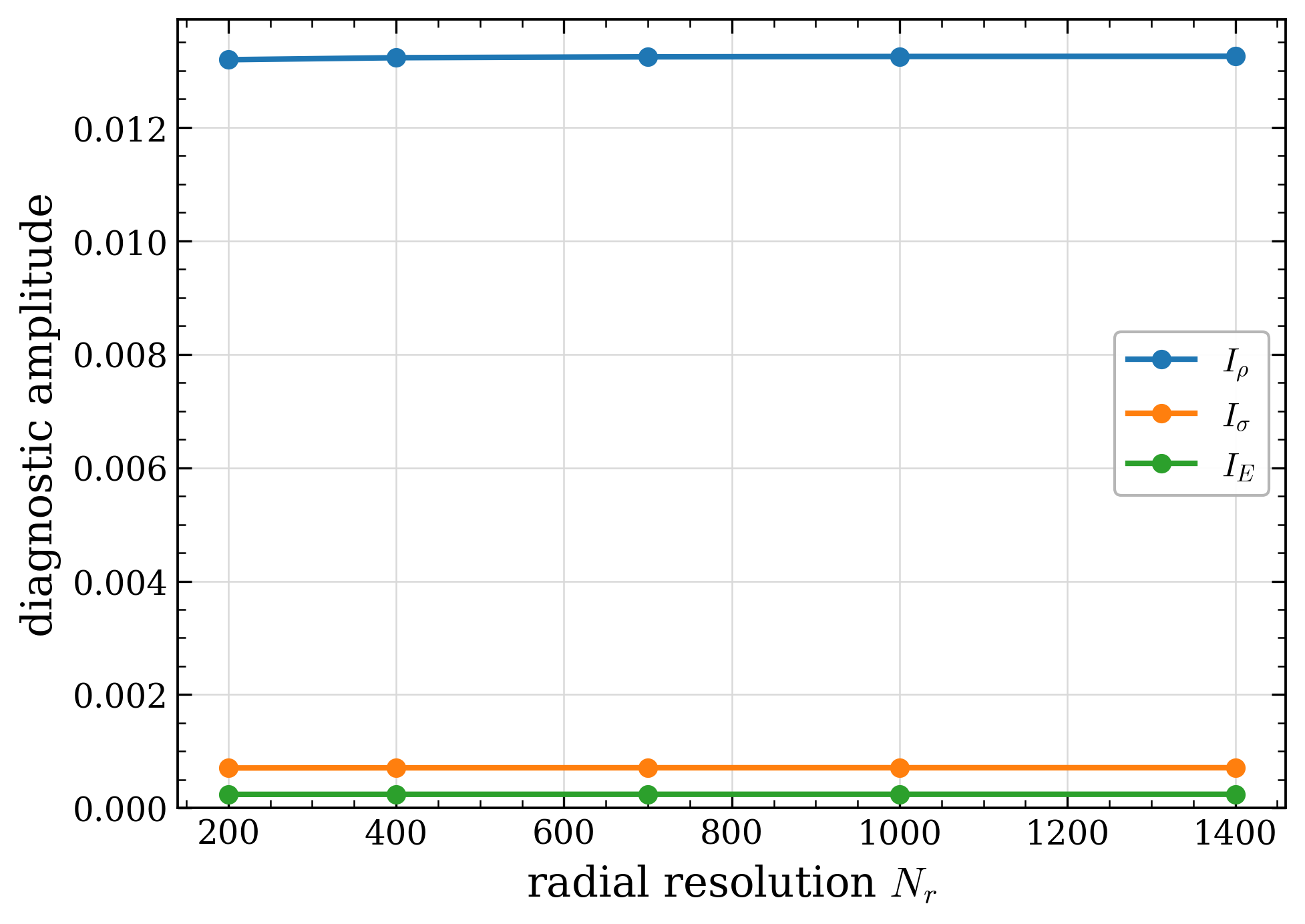}}
\caption{Robustness of the diagnostic axes. The anisotropy and amplitude sweeps rise smoothly from zero, and the radial-resolution study shows convergence at the per cent level.}\label{fig:robust}
\end{figure}

The diagnostic axes behave smoothly and predictably under controlled variation of the benchmarks, as summarised in Figure~\ref{fig:robust}. A Bianchi-I anisotropy sweep with exponents $(p-\delta,\,p,\,p+\delta)$ drives the shear and electric Weyl axes upward from zero in a smooth and monotonic manner, with $I_\sigma$ following the expected quadratic dependence on $\delta$ and reaching $10^{-2}$ at $\delta=0.2$. An amplitude sweep of the LTB bang-time profile shows every departure axis vanishing as the inhomogeneity is switched off, with matter inhomogeneity becoming dominant as the amplitude grows and $I_\rho$ reaching about $1.3\times10^{-2}$ at the largest amplitude tested. A radial-resolution study confirms numerical stability, the electric Weyl axis changing by less than one per cent as the radial resolution increases from two hundred to fourteen hundred points.

Two further tests concern the dependence on choices that form part of the construction rather than artefacts of it. Restricting the averaging domain to a central overdense region of the LTB model raises the matter inhomogeneity and electric Weyl axes substantially relative to the full domain, with $I_\rho$ rising from about $1.3\times10^{-2}$ on the full domain to about $1.2\times10^{-1}$ on the smallest central domain, although the dominant axis remains $I_\rho$ throughout. This is the expected behaviour of a domain-explicit measure, since changing the averaging domain changes the physical region being summarised, and it should be read as transparency rather than instability. A graded contamination experiment injects synthetic constraint defects of increasing amplitude into a benchmark slice. The physical classification is retained for residuals at or below the reliability tolerance, and the slice is flagged as numerically unreliable once the residual exceeds it, as shown in Table~\ref{tab:contam}. The implemented tolerance is a residual of $10^{-3}$, and the transition is visible between $10^{-3}$, which is retained, and $2\times10^{-3}$, which is flagged. These thresholds, together with the FLRW floor and the weak-departure cut, are operational choices used only to attach reproducible descriptive labels to the benchmark suite. They are not proposed as universal physical thresholds, and the continuous diagnostic vectors of Figure~\ref{fig:heatmap} and the phase-space projection remain the primary output.

\begin{table}[t]
\caption{Graded constraint contamination. Synthetic Hamiltonian and momentum residuals of increasing amplitude are injected into a benchmark slice. The implemented reliability tolerance is $10^{-3}$.}\label{tab:contam}
\begin{tabular}{@{}ccl@{}}
\toprule
Residual amplitude & Exceeds tolerance & Reported regime\\
\midrule
$0$ & no & weak anisotropy-dominated\\
$10^{-4}$ & no & weak anisotropy-dominated\\
$10^{-3}$ & no & weak anisotropy-dominated\\
$2\times10^{-3}$ & yes & numerically unreliable\\
$10^{-2}$ & yes & numerically unreliable\\
\bottomrule
\end{tabular}
\end{table}

Two further checks address the choices of normalisation and observer. We first consider the normalisation. Repeating the final-time analysis with the reference FLRW Kretschmann scale of Section~\ref{sec:axes}, in place of the unified scale $\langle\theta^4\rangle_{\mathcal D}$, changes the individual values of $I_E$ and $I_B$ but leaves the dominant axis of every benchmark unchanged. The qualitative separation reported here, therefore, does not depend on which of the two scales is used. The unified scale is adopted as the default because it removes the matter-versus-vacuum split and keeps $I_E$ and $I_B$ comparable across families.

Next, we consider observer dependence. A small radial tilt is applied to the LTB congruence, replacing the comoving observer by one carrying a peculiar velocity $v(r)=v_0\,r(1-r)$. The tilt feeds the expansion and shear axes through the velocity gradient, while the magnetic axis remains zero, because a radial boost of the spherically symmetric electric Weyl field is directed along a principal direction and preserves the purely electric character. Table~\ref{tab:tilt} reports the leading-order result. As the tilt amplitude grows from zero to $0.1$, the expansion and shear axes rise as expected, but matter inhomogeneity remains dominant, and the magnetic axis remains zero, so the classification is stable under this change of observer. A systematic study of general tilted observers and alternative foliations is left for future work, and comparisons in this paper are made within the comoving observer choice stated for each benchmark.

\begin{table}[t]
\caption{Leading-order observer-tilt sensitivity for the LTB benchmark. A radial peculiar velocity $v(r)=v_0\,r(1-r)$ is applied to the congruence. The dominant axis remains $I_\rho$ and the magnetic axis remains zero.}\label{tab:tilt}
\begin{tabular}{@{}cccccl@{}}
\toprule
$v_0$ & $I_\rho$ & $I_\theta$ & $I_\sigma$ & $I_B$ & Dominant\\
\midrule
$0$ & $1.32\times10^{-2}$ & $2.12\times10^{-3}$ & $7.1\times10^{-4}$ & $0$ & $I_\rho$\\
$0.01$ & $1.32\times10^{-2}$ & $2.20\times10^{-3}$ & $7.4\times10^{-4}$ & $0$ & $I_\rho$\\
$0.05$ & $1.32\times10^{-2}$ & $2.61\times10^{-3}$ & $8.7\times10^{-4}$ & $0$ & $I_\rho$\\
$0.1$ & $1.32\times10^{-2}$ & $3.37\times10^{-3}$ & $1.13\times10^{-3}$ & $0$ & $I_\rho$\\
\bottomrule
\end{tabular}
\end{table}

\section{Discussion}\label{sec:discussion}

The diagnostic framework proposed here should be understood as an organisational structure rather than as a new invariant construction. Its purpose is to place quantities already familiar in relativistic cosmology on a common set of non-redundant, dimensionless axes, so that different solution families can be compared as points and trajectories in a shared diagnostic space. This is particularly useful when different mechanisms of departure from FLRW behaviour coexist or compete. The clearest example is the separation of the electric and magnetic parts of the Weyl tensor. A scalar Weyl invariant alone cannot, in general, distinguish a purely electric, silent configuration from a radiative configuration, since different electric--magnetic decompositions may give the same value of $C_{abcd}C^{abcd}$. The tensor-perturbed benchmark illustrates this point directly. The magnetic Weyl axis is activated, while the non-radiative benchmarks remain at the magnetic-Weyl floor. In this sense, the electric--magnetic split supplies information that would be hidden by a scalar Weyl diagnostic alone.

The treatment of Buchert's kinematical backreaction follows the same principle. The backreaction term remains a physically meaningful explanatory quantity, especially for distinguishing variance-driven from shear-driven effects. However, its normalised magnitude is algebraically determined by the expansion-variance and shear diagnostics. Promoting it to a separate coordinate would therefore introduce a redundant axis and obscure the effective dimensionality of the diagnostic space. Reporting it as a derived quantity preserves its interpretive value while keeping the diagnostic representation non-redundant \cite{buchert2000dust,buchert2020general}.

The construction is explicitly observer-dependent, foliation-dependent, and domain-dependent. This dependence is not an artefact of the method, but a consequence of the fact that averaging and observer-based decompositions in general relativity are themselves tied to a chosen flow, hypersurface, and averaging region \cite{buchert2020general}. The framework therefore makes these choices explicit. The domain-dependence experiment illustrates this point. Restricting the LTB average to a central overdense region changes the magnitude of the matter-inhomogeneity and electric-Weyl diagnostics, while preserving the dominant physical interpretation. The leading-order observer-tilt test similarly shows that, for the radial LTB tilt considered here, the dominant axis remains stable even though the expansion and shear diagnostics respond to the change of congruence. These tests do not exhaust the possible observer and foliation choices, but they demonstrate how such dependence can be reported quantitatively rather than hidden.

The reliability indicators $I_{\mathcal H}$ and $I_{\mathcal M}$ play a complementary role. In the analytic and perturbative benchmarks considered here, they remain at round-off because the solutions are either exact or constructed to satisfy the relevant linearised constraints. In a numerical-relativity setting, by contrast, these quantities would encode the actual Hamiltonian and momentum constraint residuals of the simulation. The contamination experiment demonstrates how this information can be incorporated without mixing numerical reliability with physical classification. Below the chosen tolerance the physical diagnostic is retained, while above, it the state is flagged as unreliable. This separation is important for applications to numerical-relativity cosmology, where physical structure and numerical error must be distinguished carefully \cite{macpherson2017nr,aurrekoetxea2025review}.

Several limitations should be noted. The benchmark suite is analytic or semi-analytic, and the scalar and tensor perturbative examples are treated at linear order. No fully nonlinear numerical-relativity slice is analysed in the present work. The thresholds used to label FLRW-like, weakly departed, or unreliable states are operational choices rather than universal physical constants. The continuous diagnostic vector should therefore be regarded as the primary result, with the labels serving only as reproducible summaries. The observer-tilt test is also deliberately limited to a leading-order radial tilt of the LTB congruence. More general tilted observers and alternative foliations may produce different quantitative diagnostics and should be studied separately.

A natural next exact benchmark is a quasi-spherical Szekeres model. Such a model would test the framework beyond spherical symmetry while remaining an exact inhomogeneous cosmological solution. For the comoving congruence it is expected to remain purely electric, so its role would be to test non-symmetric electric-Weyl structure rather than to activate the magnetic axis. As with the Bianchi-I and LTB benchmarks used here, it should be incorporated only after its curvature expressions have been verified against metric-derived invariants \cite{szekeres1975,vrbasvitek2014}. Other extensions include tilted-observer studies, applications to full numerical-relativity cosmological simulations, and the use of the diagnostic space for automated regime discovery in large ensembles of relativistic solutions \cite{bittencourt2014,stephani2003}.

\section{Conclusion}\label{sec:conclusion}

In this paper, we have introduced a geometric diagnostic phase space for comparing cosmological dynamics in Einstein's field equations. The construction organises standard kinematic, curvature, and constraint quantities into a compact set of non-redundant diagnostics that distinguish different mechanisms of departure from FLRW behaviour. Across the benchmark suite considered here, the framework separates homogeneous anisotropy, vacuum anisotropy, matter inhomogeneity, electric-Weyl structure, magnetic and radiative structure, and numerical unreliability. The resulting diagnostic vectors place the six benchmark families in distinct regions of the phase space, with the magnetic-Weyl axis activated only by the tensor-perturbed benchmark.

A central feature of the construction is that Buchert's kinematical backreaction is retained as a derived explanatory quantity rather than promoted to a separate coordinate. This avoids introducing an algebraically redundant axis while preserving the ability to distinguish variance-driven from shear-driven backreaction. The analytic curvature expressions used for the exact benchmarks have been verified symbolically against metric-derived Weyl invariants, and the perturbative benchmarks have been checked for the relevant constraint consistency. Robustness tests further show that the qualitative classification is stable under changes in perturbation amplitude, spatial resolution, averaging domain, reliability threshold, curvature normalisation, and a leading-order observer tilt.

The framework is not intended as a foliation-independent invariant classification of spacetime. Rather, it provides an observer- and domain-explicit diagnostic for comparing analytic, perturbative, and numerical cosmological solution families on a common geometric footing. In this form, it offers a reproducible basis for studying how relativistic cosmological models depart from the FLRW reference and for applying the same diagnostic structure to future numerical-relativity simulations and broader families of exact inhomogeneous solutions.

\section*{Acknowledgments}
The author acknowledges the computational resources provided by the Centre for Visual Computing and Intelligent Systems at the University of Bradford.

\section*{Data availability}\label{sec:data}
The computer source code, exported tables, figures, metadata, and result archives behind the analyses is available in the GitHub repository at:\\ 
\url{https://github.com/ugail/Phase-Space-Structure-for-Einstein-Field-Equations} \\
The repository holds the main computational notebook, the symbolic verification routines, the benchmark result tables, and all figure-generation code. The repository is designed to reproduce all reported numerical results, figures, and tables from a clean run.

\section*{Funding}\label{fundning_}
No funding was received for this work. 

\subsection*{Competing interests}\label{conpeting_}
The author declares no competing interests.

\begin{appendices}

\section{Mathematical and numerical details}\label{app:details}

This appendix records the closed-form expressions, the symbolic verification, and the numerical settings in sufficient detail to reproduce the analyses.

The Bianchi-I Weyl invariant is obtained from the metric $\mathrm{d}s^2=-\mathrm{d}t^2+\sum_i t^{2p_i}\,(\mathrm{d}x^i)^2$ by forming the kinematic quantities $H_i=p_i/t$, $\theta=\sum_i H_i$, and $\sigma^2=\tfrac12\sum_i(H_i-\theta/3)^2$, together with the Kretschmann, Ricci, and Ricci-scalar combinations that yield $C_{abcd}C^{abcd}$. The closed form obtained equals the invariant computed directly from the metric as an exact symbolic identity, vanishes when the exponents are equal, and returns the vacuum Kasner value when the Kasner conditions hold. The LTB Weyl invariant is given in closed form by Equation~\eqref{eq:ltbweyl}, and for the parabolic dust solution with $R=(9M/2)^{1/3}r\,(t-t_B(r))^{2/3}$ it agrees with the metric-derived invariant to machine precision across a grid in $(t,r)$, with the field equation $\dot R^2=2M/R$ satisfied identically. The electric and magnetic Weyl tensors of the gravitational-wave benchmark were computed from the linearised transverse-traceless metric on a radiation background, and the electric and magnetic invariants were averaged over one wave period. For a plane wave in flat space $E_{ab}E^{ab}=B_{ab}B^{ab}$ exactly, whereas on the expanding background the finite ratio of wavelength to Hubble scale leaves them comparable rather than equal, with $|I_E-I_B|/\max(I_E,I_B)\approx0.13$ at the wavenumber used, decreasing as the mode becomes more sub-horizon. The Weyl scalar $C_{abcd}C^{abcd}=8(E_{ab}E^{ab}-B_{ab}B^{ab})$ is therefore suppressed relative to either Weyl part while $B_{ab}B^{ab}>0$ remains, which is what activates the magnetic axis.

The scalar perturbation benchmark is constructed in the longitudinal gauge for a matter-dominated growing mode with $\Phi$ independent of time. The density and velocity perturbations are fixed by the linearised Hamiltonian and momentum constraints, with spatial derivatives evaluated spectrally on a periodic grid, so that the constraint residuals lie at round-off and the electric Weyl part follows from the trace-free Hessian of $\Phi$ in the proper frame. The gravitational-wave benchmark uses a single transverse-traceless mode on a radiation background, and its electric and magnetic invariants are averaged over one wave period so that the oscillating mode yields smooth and positive axis values.

The numerical settings are as follows. The homogeneous benchmarks are evaluated on a time grid of eighty points. The LTB model uses a radial grid that ranges from two hundred to fourteen hundred points for the resolution study and is fixed at seven hundred points elsewhere, with derivatives taken by finite differences and domain averages weighted by the proper radial volume element. The perturbative benchmarks use a periodic spatial grid with spectral derivatives. The regulator $\epsilon$ is set to a small fixed value that affects only the homogeneous limit. The reliability tolerance is $10^{-3}$, the FLRW floor below which a state is called FLRW-like is $10^{-4}$, and a state whose dominant axis lies below $10^{-2}$ is labelled weak.

\end{appendices}


\begin{thebibliography}{99}

\bibitem{buchert2000dust}
T.~Buchert, ``On average properties of inhomogeneous fluids in general relativity: dust cosmologies,'' \emph{General Relativity and Gravitation}, vol.~32, pp.~105--125, 2000. \url{https://doi.org/10.1023/A:1001800617177}

\bibitem{buchert2001perfect}
T.~Buchert, ``On average properties of inhomogeneous fluids in general relativity: perfect fluid cosmologies,'' \emph{General Relativity and Gravitation}, vol.~33, pp.~1381--1405, 2001. \url{https://doi.org/10.1023/A:1012061725841}

\bibitem{bck2011}
K.~Bolejko, M.-N.~C\'el\'erier, and A.~Krasi\'nski, ``Inhomogeneous cosmological models: exact solutions and their applications,'' \emph{Classical and Quantum Gravity}, vol.~28, art.~164002, 2011. \url{https://doi.org/10.1088/0264-9381/28/16/164002}

\bibitem{macpherson2017nr}
H.~J.~Macpherson, P.~D.~Lasky, and D.~J.~Price, ``Inhomogeneous cosmology with numerical relativity,'' \emph{Physical Review D}, vol.~95, art.~064028, 2017. \url{https://doi.org/10.1103/PhysRevD.95.064028}

\bibitem{aurrekoetxea2025review}
J.~C.~Aurrekoetxea, K.~Clough, and E.~A.~Lim, ``Cosmology using numerical relativity,'' \emph{Living Reviews in Relativity}, vol.~28, art.~5, 2025. \url{https://doi.org/10.1007/s41114-025-00058-z}

\bibitem{ugail2026qnn}
H.~Ugail and N.~Howard, ``Symmetry-organised complexity in quantum neural networks,'' \emph{Symmetry}, vol.~18, no.~6, art.~912, 2026. \url{https://doi.org/10.3390/sym18060912}

\bibitem{ellisvanelst1999}
G.~F.~R.~Ellis and H.~van Elst, ``Cosmological models (Carg\`ese lectures 1998),'' in \emph{Theoretical and Observational Cosmology}, NATO Science Series C, vol.~541, pp.~1--116, Kluwer, Dordrecht, 1999. \url{https://arxiv.org/abs/gr-qc/9812046}

\bibitem{emm2012}
G.~F.~R.~Ellis, R.~Maartens, and M.~A.~H.~MacCallum, \emph{Relativistic Cosmology}. Cambridge University Press, Cambridge, 2012.

\bibitem{wald1984}
R.~M.~Wald, \emph{General Relativity}. University of Chicago Press, Chicago, 1984.

\bibitem{costanatario2014}
L.~F.~O.~Costa and J.~Nat\'ario, ``Gravito-electromagnetic analogies,'' \emph{General Relativity and Gravitation}, vol.~46, art.~1792, 2014. \url{https://doi.org/10.1007/s10714-014-1792-1}

\bibitem{clifton2017}
T.~Clifton, D.~Gregoris, and K.~Rosquist, ``The magnetic part of the Weyl tensor, and the expansion of discrete universes,'' \emph{General Relativity and Gravitation}, vol.~49, art.~30, 2017. \url{https://doi.org/10.1007/s10714-017-2192-0}

\bibitem{kasner1921}
E.~Kasner, ``Geometrical theorems on Einstein's cosmological equations,'' \emph{American Journal of Mathematics}, vol.~43, no.~4, pp.~217--221, 1921. \url{https://doi.org/10.2307/2370192}

\bibitem{wainwrightkrasinski2008}
J.~Wainwright and A.~Krasi\'nski, ``Republication of: geometrical theorems on Einstein's cosmological equations,'' \emph{General Relativity and Gravitation}, vol.~40, pp.~865--876, 2008. \url{https://doi.org/10.1007/s10714-007-0574-4}

\bibitem{misner1969}
C.~W.~Misner, ``Mixmaster universe,'' \emph{Physical Review Letters}, vol.~22, pp.~1071--1074, 1969. \url{https://doi.org/10.1103/PhysRevLett.22.1071}

\bibitem{lemaitre1933}
G.~Lema\^itre, ``L'univers en expansion,'' \emph{Annales de la Soci\'et\'e Scientifique de Bruxelles A}, vol.~53, pp.~51--85, 1933.

\bibitem{tolman1934}
R.~C.~Tolman, ``Effect of inhomogeneity on cosmological models,'' \emph{Proceedings of the National Academy of Sciences}, vol.~20, no.~3, pp.~169--176, 1934. \url{https://doi.org/10.1073/pnas.20.3.169}

\bibitem{bondi1947}
H.~Bondi, ``Spherically symmetrical models in general relativity,'' \emph{Monthly Notices of the Royal Astronomical Society}, vol.~107, no.~5--6, pp.~410--425, 1947. \url{https://doi.org/10.1093/mnras/107.5-6.410}

\bibitem{krasinski1997}
A.~Krasi\'nski, \emph{Inhomogeneous Cosmological Models}. Cambridge University Press, Cambridge, 1997.

\bibitem{bardeen1980}
J.~M.~Bardeen, ``Gauge-invariant cosmological perturbations,'' \emph{Physical Review D}, vol.~22, no.~8, pp.~1882--1905, 1980. \url{https://doi.org/10.1103/PhysRevD.22.1882}

\bibitem{kodamasasaki1984}
H.~Kodama and M.~Sasaki, ``Cosmological perturbation theory,'' \emph{Progress of Theoretical Physics Supplement}, vol.~78, pp.~1--166, 1984. \url{https://doi.org/10.1143/PTPS.78.1}

\bibitem{mfb1992}
V.~F.~Mukhanov, H.~A.~Feldman, and R.~H.~Brandenberger, ``Theory of cosmological perturbations,'' \emph{Physics Reports}, vol.~215, no.~5--6, pp.~203--333, 1992. \url{https://doi.org/10.1016/0370-1573(92)90044-Z}

\bibitem{buchert2020general}
T.~Buchert, P.~Mourier, and X.~Roy, ``On average properties of inhomogeneous fluids in general relativity III: general fluid cosmologies,'' \emph{General Relativity and Gravitation}, vol.~52, art.~27, 2020. \url{https://doi.org/10.1007/s10714-020-02670-6}

\bibitem{szekeres1975}
P.~Szekeres, ``A class of inhomogeneous cosmological models,'' \emph{Communications in Mathematical Physics}, vol.~41, pp.~55--64, 1975. \url{https://doi.org/10.1007/BF01608547}

\bibitem{vrbasvitek2014}
D.~Vrba and O.~Sv\'itek, ``Modelling inhomogeneity in Szekeres spacetime,'' \emph{General Relativity and Gravitation}, vol.~46, art.~1808, 2014. \url{https://doi.org/10.1007/s10714-014-1808-x}

\bibitem{bittencourt2014}
E.~Bittencourt, J.~M.~Salim, and G.~B.~dos~Santos, ``Magnetic fields and the Weyl tensor in the early universe,'' \emph{General Relativity and Gravitation}, vol.~46, art.~1790, 2014. \url{https://doi.org/10.1007/s10714-014-1790-3}

\bibitem{stephani2003}
H.~Stephani, D.~Kramer, M.~A.~H.~MacCallum, C.~Hoenselaers, and E.~Herlt, \emph{Exact Solutions of Einstein's Field Equations}, 2nd~ed. Cambridge University Press, Cambridge, 2003.

\end{thebibliography}
\end{document}